\documentclass[preprint,prd,amsmath,amssymb,nofootinbib,tightenlines,floatfix]{revtex4}
\usepackage[dvips]{graphics}
\usepackage{epsfig}
\usepackage{latexsym}
\newcommand{\lsim}{\buildrel < \over {_\sim}}
\newcommand{\gsim}{\buildrel > \over {_\sim}}

\newcommand{\be}{\begin{equation}}
\newcommand{\ee}{\end{equation}}
\newcommand{\bea}{\begin{eqnarray}}
\newcommand{\eea}{\end{eqnarray}}
\newcommand{\ba}{\begin{array}}
\newcommand{\ea}{\end{array}}

\begin{document}

\preprint{Caltech MAP-310}
\preprint{CALT-68-2588}

\title{\Large Neutrino Mass Implications for Muon Decay Parameters}

\author{Rebecca J. Erwin\footnote{
	Electronic address: rjerwin@caltech.edu}
}
\affiliation{California Institute of Technology, Pasadena, CA 91125}

\author{Jennifer Kile\footnote{
	Electronic address: jenkile@theory.caltech.edu}
}
\affiliation{California Institute of Technology, Pasadena, CA 91125}

\author{Michael J. Ramsey-Musolf\footnote{
	Electronic address: mjrm@caltech.edu}}
\affiliation{California Institute of Technology, Pasadena, CA 91125}

\author{Peng Wang\footnote{
	Electronic address: pengw@theory.caltech.edu}
}
\affiliation{California Institute of Technology, Pasadena, CA 91125}


%

\begin{abstract}

We use the scale of neutrino mass to derive model-independent naturalness constraints on possible contributions to muon decay Michel parameters from new physics above the electroweak symmetry-breaking scale. Focusing on Dirac neutrinos, we obtain a complete basis of effective dimension four and dimension six operators that are invariant under the gauge symmetry of the Standard Model and that contribute to both muon decay and neutrino mass. We show that -- in the absence of fine tuning -- the most stringent bounds on chirality-changing operators relevant to muon decay arise from one-loop contributions to neutrino mass. The bounds we obtain on their contributions to the Michel parameters are four or more orders of magnitude stronger than bounds  previously obtained in the literature.  We also show that there exist chirality-changing operators that contribute to muon decay but whose flavor structure allows them to evade neutrino mass naturalness bounds. We discuss the implications of our analysis for the interpretation of muon decay experiments.

\end{abstract}

\maketitle


\section{Introduction} 

Precision studies of muon decay continue to play an important role in testing the Standard Model (SM) and searching for physics beyond it. In the gauge sector of the SM, the Fermi constant $G_\mu$ that characterizes the strength of the low-energy, four-lepton $\mu$-decay operator is determined from the $\mu$ lifetime and gives one of the three most precisely-known inputs into the theory. Analyses of the spectral shape, angular distribution, and polarization of the decay electrons (or positrons) probe for contributions from operators that deviate from the $(V-A)\otimes (V-A)$ structure of the SM decay operator. In the absence of time-reversal (T) violating interactions, there exist seven independent parameters -- the so-called Michel parameters\cite{Michel,Kinoshita} -- that characterize the final state charged leptons: two ($\rho$, $\eta$) that describe the spatially isotropic component of the lepton spectrum; two ($\xi$, $\delta$) that characterize the spatially anisotropic distribution; and three additional quantities ($\xi^\prime$, $\xi^{\prime\prime}$, $\eta^{\prime\prime}$) that are needed to describe the lepton's transverse and longitudinal polarization\footnote{The parameters $\eta$ and $\eta^{\prime\prime}$ are alternately written in terms of the independent parameters  $\alpha/A$ and $\beta/A$.}. Two additional parameters ($\alpha^\prime/A$, $\beta^\prime/A$) characterize a T-odd correlation between the final state lepton spin and momenta with the muon polarization: ${\hat S}_e\cdot {\hat k}_e\times{\hat S}_\mu$. 

Recently, new experimental efforts have been devoted to more precise determinations of these parameters. The TWIST Collaboration has measured $\rho$ and $\delta$ at TRIUMF\cite{Musser:2004zw,Gaponenko:2004mi}, improving the uncertainty over previously reported values by factors of $\sim 2.5$ and $\sim 3$, respectively. An experiment to measure the transverse positron polarization has been carried out at the Paul Scherrer Institute (PSI), leading to similar improvements in sensitivity over the results of earlier measurements\cite{Danneberg:2005xv}. A new determination of $P_\mu\xi$ with a similar degree of improved precision is expected from the TWIST Collaboration, and one anticipates additional reductions in the uncertainties in $\rho$ and $\delta$\cite{tribble}. 

At present, there exists no evidence for deviations from SM predictions for the Michel parameters (MPs). It is interesting, nevertheless, to ask what constraints these new measurements can provide on possible contributions from physics beyond the SM. It has been conventional to characterize these contributions in terms of a set of ten four-fermion operators
\be
\label{eq:leff0}
{\cal L}^{\mu-\rm decay} = - \frac{4 G_\mu}{\sqrt{2}}\ \sum_{\gamma,\, \epsilon,\, \mu} \ g^\gamma_{\epsilon\mu}\, 
\ {\bar e}_\epsilon \Gamma^\gamma \nu {\bar\nu} \Gamma_\gamma \mu_\mu
\ee
where the sum runs over Dirac matrices $\Gamma^\gamma= 1$ (S), $\gamma^\alpha$ (V), and $\sigma^{\alpha\beta}/\sqrt{2}$ (T) and the subscripts $\mu$ and $\epsilon$  denote the chirality ($R$,$L$) of the muon and final state lepton, respectively\footnote{The normalization of the tensor terms corresponds to the convention adopted in Ref.~\cite{Scheck}. We do not specify the neutrino flavors in Eq. (\ref{eq:leff0}) since the $\mu$-decay experiments do not observe the final state neutrinos. }. In the SM, one has $g^V_{LL}=1$ and all other $g^\gamma_{\epsilon\mu}=0$. A recent, global analysis by Gagliardi, Tribble, and Williams \cite{Gagliardi:2005fg} give the present experimental bounds on the $g^\gamma_{\epsilon\mu}$ that include the impact of the latest TRIUMF and PSI measurements.

Theoretically, the $g^\gamma_{\epsilon\mu}$ can be generated in different scenarios for physics beyond the SM. The most commonly cited illustration is the minimal left-right symmetric model that gives rise to non-zero $g^V_{RR}$, $g^V_{RL}$, and $g^V_{LR}$. From a model-independent standpoint,  the authors of Ref.~\cite{Prezeau:2004md} recently observed that the operators in Eq.~(\ref{eq:leff0}) having different chiralities for the muon and final state charged lepton will also contribute to the neutrino mass matrix $m_\nu^{AB}$ through radiative corrections.  Consequently, one expects that the present  upper bounds on $m_\nu$ should imply bounds on the magnitudes of the $g^\gamma_{\epsilon\mu}$. The authors of Ref.~\cite{Prezeau:2004md} argued that the most stringent limits arise from two-loop contributions because the one-loop contributions are suppressed by three powers of the tiny, charged lepton Yukawa couplings. The two-loop constraints are nonetheless stronger than the present bounds give in Ref.~\cite{Gagliardi:2005fg} and could become even more so with the advent of future terrestrial and cosmological probes of the neutrino mass scale.

In this paper, we present the results of a follow-up analysis of $m_\nu$ constraints on the $\mu$-decay parameters, motivated by the observations of Ref.~\cite{Prezeau:2004md} and the new experimental developments in the field. Our study follows the approach of Ref.~\cite{Bell:2005kz,Davidson:2005cs} used recently in deriving model-independent naturalness bounds on neutrino magnetic moments implied by the scale of $m_\nu$. For concreteness, we work with an effective theory that is valid below a scale $\Lambda$ lying above the weak scale $v\approx 246$ GeV and that contains SU(2)$_L\times$U(1)$_Y$-invariant 
operators built from Standard Model fields plus right-handed (RH) Dirac neutrinos\footnote{We defer a study of the effective theory containing Majorana neutrinos to a subsequent publication.}. We consider all relevant operators up to dimension $n=6$ that could be generated by physics above the scale $\Lambda$. For simplicity, we restrict our attention to two generations of lepton doublets and RH neutrinos. Extending the analysis to include a third generation  increases the number of relevant operators but does not change the substantive conclusions. While the spirit of our work is similar to that of  Ref.~\cite{Prezeau:2004md}, the specifics of our analysis and conclusions differ in several respects:

\begin{itemize}

\item[i)] The effective theory that we adopt allows us to compute contributions to $m_\nu$ from scales lying between the weak scale $v$ and the scale of new physics $\Lambda$. In contrast, the authors of Ref.~\cite{Prezeau:2004md} used a Fierz transformed version of ${\cal L}^{\mu-\rm decay} $ in Eq. (\ref{eq:leff0}), which is not  invariant under the SM gauge group  and, therefore, should  be used to analyze only contributions below the weak scale. 

\item[ii)]  We show that for the two flavor case the operators in ${\cal L}^{\mu-\rm decay} $ proportional to $g^{S,T}_{LR}$ and $g^{S,T}_{RL}$  arise from twelve independent dimension $n=6$  gauge-invariant four-fermion operators, while those containing $g^V_{LR}$ and $g^V_{RL}$ are generated by four independent $n=6$ operators that contain two fermions and two Higgs scalars. 

\item[iii)] While the operators that contribute to $\mu$-decay have dimension $n=6$ or higher, the lowest dimension neutrino mass operator occurs at $n=4$. The authors of Ref.~\cite{Prezeau:2004md} used dimensional regularization (DR) to analyze the mixing between the $n=6$ $\mu$-decay and neutrino mass operators, but did not consider mixing with the $n=4$ operator that cannot be determined with DR. We derive order-of-magnitude constraints on the $n=6$ operator coefficients  implied by this mixing, which depends only linearly on the lepton Yukawa couplings and which gives the dominant constraints for $\Lambda \gg v$.

\item[iv)] For $\Lambda$ not too different from $v$, constraints associated with mixing among the $n=6$ operators can, in principle, be comparable to those arising from contributions to the $n=4$ mass operator. We carry out a complete, one-loop analysis of this mixing and show that only the  neutrino magnetic moment and two-fermion/two-Higgs operators mix with the $n=6$ neutrino mass operator to linear order in the lepton Yukawa couplings. We derive the resulting bounds on the $g^V_{LR,RL}$ that follow from this mixing and find that they are comparable to those associated with induced $n=4$ mass operator for $\Lambda \gsim v$.

\item[v)] From the mixing with the $n=4$ and $n=6$ mass operators, we find that the bounds on the $|g^{V}_{LR,RL}|$ are two or more orders of magnitude stronger than obtained in Ref.~\cite{Prezeau:2004md} and at least three orders of magnitude below the experimental limits given in Ref.~\cite{Gagliardi:2005fg}. 

\item[vi)] The neutrino mass implications for the couplings $g^{S,T}_{LR,RL}$ are more subtle. Of the twelve independent four-fermion operators that contribute to these couplings, only eight are directly constrained by the scale of neutrino mass and naturalness considerations. Their contributions to the $g^{S,T}_{LR,RL}$ are generally $\sim 10^4$ times smaller than the present experimental bounds, and $\sim 10^3$ times smaller than obtained in the analysis of Ref.~\cite{Prezeau:2004md}. 
We show, however,  that the flavor structure of the remaining four operators allows them to evade such constraints. While from a theoretical perspective one might not expect their contributions to be substantially larger than those from the constrained operators, experimental efforts to determine the $g^{S,T}_{LR,RL}$ remain a worthwhile endeavor.

\end{itemize}

A summary of our results is given in Table \ref{tab:gconstraints}. In the remainder of the paper we give the details of our analysis. In Section \ref{sec:basis}, we write down the complete set of independent operators through $n=6$ that contribute to $m_\nu^{AB}$ and/or $\mu$-decay. Section \ref{sec:mixing} gives our analysis of operator mixing, while in Section 
\ref{sec:mnuconstraints} we discuss the resulting constraints on the $g_{LR,RL}^\gamma$ that follow from this analysis and the present upper bounds on the neutrino mass scale. We summarize in Section \ref{sec:conclusions}. The Appendix discusses the sensitivity of the Michel parameters to the neutrino magnetic moment operators.

\begin{table}
\caption{Constraints on $\mu$-decay couplings $g^\gamma_{\epsilon\mu}$. The first eight rows give naturalness bounds  in units of $(v/\Lambda)^2\times (m_\nu/1\, {\rm eV})$ on contributions from $n=6$ muon decay operators (defined in Section \ref{sec:basis} below) based on one-loop mixing with the $n=4$ neutrino mass operators. The ninth row gives upper bounds derived from a recent global analysis of Ref.~\cite{Gagliardi:2005fg}, while the last row gives estimated bounds from Ref.~\cite{Prezeau:2004md} derived from two-loop mixing of $n=6$ muon decay and mass operators. A \lq\lq -" indicates that the operator does not contribute to the given $g^\gamma_{\epsilon\mu}$, while \lq\lq None" indicates that the operator gives a contribution unconstrained by neutrino mass. The subscript $D$ runs over the two generations of RH Dirac neutrinos. }
\label{tab:gconstraints}
\begin{tabular}{@{}lcccccc @{}}
\\
Source & $|g^S_{LR}|$  & $|g^T_{LR}|$ & $|g^S_{RL}|$ & $|g^T_{RL}|$ & $|g^V_{LR}|$ & $|g^V_{RL}|$
\\
\hline
\\

${\cal O}^{(6)}_{F,\, 122D}$ & $4\times 10^{-7}$ & $2\times 10^{-7}$ & - & - & - & - \\ 
${\cal O}^{(6)}_{F,\, 212D}$ & $4\times 10^{-7}$ & - & - & - & - & - \\ 
${\cal O}^{(6)}_{F,\, 112D}$ & None & None & - & - & - & - \\  
${\cal O}^{(6)}_{F,\, 211D}$  & - & -&  $8\times 10^{-5}$ & $4\times 10^{-5}$ & - & - \\ 
${\cal O}^{(6)}_{F,\, 121D}$ & - & - & $8\times 10^{-5}$ & - & - & - \\ 
${\cal O}^{(6)}_{F,\, 221D}$ & - & - &  None & None & - & - \\
${\cal O}^{(6)}_{{\tilde V},\, 2D}$ & - & - & - & - & $8\times 10^{-7}$ & - \\ 
${\cal O}^{(6)}_{{\tilde V},\, 1D}$ & - & - & - & - & - & $2 \times 10^{-4}$  \\ \\ \hline

Global~\cite{Gagliardi:2005fg} & 0.088 & 0.025 & 0.417 & 0.104 & 0.036 & 0.104\\ 
Two-loop~\cite{Prezeau:2004md} & $10^{-4}$ & $10^{-4}$ & $10^{-2}$ & $10^{-2}$ & $10^{-4}$ & $10^{-2}$ \\
\hline
\end{tabular}
\end{table}

\section{Operator Basis}
\label{sec:basis}

To set notation, we follow Ref. \cite{Bell:2005kz} and consider the effective Lagrangian
\begin{equation}
\label{eq:leff1}
{\cal L}_{\rm eff} = \sum_{n,j} \frac{C_j^n(\mu)}{\Lambda^{n-4}}\, {\cal O}_j^{(n)}(\mu) \  + {\rm h.c.}
\end{equation}
where $\mu$ is the renormalization scale, $n\geq 4$ is the operator dimension, and $j$ is an index running over all independent operators of a given dimension. The lowest dimension neutrino mass operator is
\be
{\cal O}^{(4)}_{M,\, AD} = {\bar L}^A {\tilde\phi} \nu_R^D
\ee
where $L^A$ is the left-handed (LH) lepton doublet for generation $A$, $\nu_R^D$ is a RH neutrino for generation $D$ and ${\tilde\phi}=i\tau_2\phi^\ast$ with $\phi$ being the Higgs doublet field. After spontaneous symmetry breaking, one has
\be
\phi\rightarrow\left(\begin{array}{c} 0 \\ v/\sqrt{2} \end{array}\right)
\ee
so that 
\begin{eqnarray}
\nonumber
C_{M,\, AD}^4 {\cal O}^{(4)}_{M,\, AD}&\rightarrow & -m_\nu^{AD} {\bar\nu}_L^A \nu_R^D\\
\label{eq:neq4mass}
m_\nu^{AD} &=& -C^4_{M,\, AD}\, v/\sqrt{2}\ \ \ .
\end{eqnarray} 
The other $n=4$ operators are those of the SM and we do not write them down explicitly here.

For the case of Dirac neutrinos that we consider here, there exist no gauge-invariant $n=5$ operators. In considering those with dimension six, it is useful to group them according to the number of fermion, Higgs, and gauge boson fields that enter:

\vskip 0.25in
\noindent {\em Four fermion:}
\vskip 0.25in

\begin{eqnarray}
\nonumber
\bar{L}\gamma^{\mu}L\bar{L}\gamma_{\mu}L \\
\nonumber
\bar{\ell}_{R}\gamma^{\mu}\ell_{R}\bar{\ell}_{R}\gamma_{\mu}\ell_{R} \\
\nonumber
\bar{\ell}_{R}\gamma^{\mu}\ell_{R}\bar{\nu_{R}}\gamma_{\mu}\nu_{R} \\
\nonumber
\bar{\nu_{R}}\gamma^{\mu}\nu_{R}\bar{\nu_{R}}\gamma_{\mu}\nu_{R} \\
\nonumber
\bar{L}\ell_{R}\bar{\ell}_{R}L\\
\nonumber
\bar{L}\nu_{R}\bar{\nu_{R}}L\\
\nonumber
\epsilon^{ij}\bar{L}_{i}\ell_{R}\bar{L}_{j}\nu_{R}
\end{eqnarray}
Here $\ell_R$ is the right-handed charged lepton field. 
Several of the operators appearing in this list can contribute to $\mu$-decay, but only the last one can also contribute to $m_\nu^{AD}$ through radiative corrections. Including flavor indices, we refer to this operator as
\be
{\cal O}^{(6)}_{F,\, ABCD} = \epsilon^{ij}\bar{L}_{i}^A\ell_{R}^C\bar{L}_{j}^B\nu_{R}^D
\ee
where the indices $i,j$ refer to the weak isospin components of the LH doublet fields and $\epsilon^{12}=-\epsilon^{21}=1$.

\begin{figure}[h]
\epsfxsize=5in
\centerline{\epsfig{figure=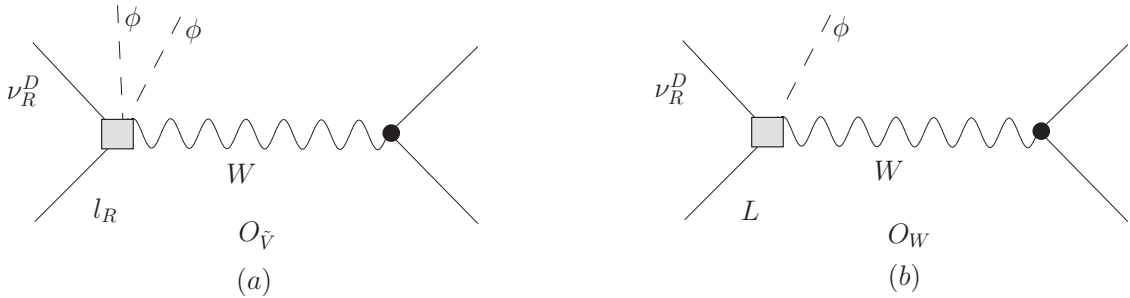,width=6.in}} \caption{{\em  Contributions from the operators (a) ${\cal O}^{(6)}_{{\tilde V},\, AD}$ and (b) ${\cal O}^{(6)}_{W,\, AD}$ (denoted by the shaded box) to the amplitude for $\mu$-decay. Solid, dashed, and wavy lines denote fermions, Higgs scalars, and gauge bosons, respectively. After SSB, the neutral Higgs field is replaced by its vev, yielding a four-fermion $\mu$-decay amplitude.}} \label{fig:mutree}
\end{figure}

\vskip 0.25in
\noindent {\em Fermion-Higgs:}
\vskip 0.25in

\begin{eqnarray}
\nonumber
i(\bar{L}^A\gamma^{\mu}L^B)(\phi^{+}D_{\mu}\phi) \\
\nonumber
i(\bar{L}^A\gamma^{\mu}\tau^{a}L^B)(\phi^{+}\tau^{a}D_{\mu}\phi) \\
\label{eq:fermionhiggs}
i(\bar{\ell}_{R}^A\gamma^{\mu}\ell_{R}^B)(\phi^{+}D_{\mu}\phi)\\
\nonumber
i(\bar{\nu}_{R}^A\gamma^{\mu}\nu_{R}^B)(\phi^{+}D_{\mu}\phi)\\
\nonumber
i(\bar{\ell}_{R}^A\gamma^{\mu}\nu_{R}^B)(\phi^{+}D_{\mu}\widetilde{\phi})
\end{eqnarray}
Neither of the first two operators in the list (\ref{eq:fermionhiggs}) can contribute significantly to $m_\nu^{AD}$  since they contain no RH neutrino fields. Any loop graph through which they radiatively induce $m_\nu^{AD}$ would have to contain operators that contain both LH and RH fields, such as ${\cal O}^{(4)}_{M,\, AB}$ or other $n=6$ operators. In either case, the resulting constraints on the operator coefficients will be weak. For similar reasons, the third and fourth  operators cannot contribute substantially because they contain an even number of neutrino fields having the same chirality and since the neutrino mass operator contains one LH and one RH neutrino field. Only the last operator 
\be
{\cal O}^{(6)}_{{\tilde V},\, AD}  \equiv i(\bar{\ell}_{R}^A\gamma^{\mu}\nu_{R}^D)(\phi^{+}D_{\mu}\widetilde{\phi})
\ee
can contribute signficantly to $m_\nu$ since it contains a single RH neutrino. It also contributes to the $\mu$-decay amplitude after SSB via the graph of Fig. \ref{fig:mutree}a  since the covariant derivative $D_\mu$ contains charged $W$-boson fields. We also write down the $n=6$ neutrino mass operators
\be
{\cal O}^{(6)}_{M,\, AD} = (\bar{L}^A\widetilde{\phi}\nu_{R}^D)(\phi^{+}\phi)
\ee
as well as the charged lepton mass operator $(\bar{L}\phi \ell_{R})(\phi^{+}\phi)$ that we do not use in the present analysis.

\vskip 0.25in
\noindent {\em Fermion-Higgs-Gauge:}
\vskip 0.25in

\begin{eqnarray}
\nonumber
\bar{L}\tau^{a}\gamma^{\mu}D^{\nu}LW_{\mu\nu}^{a}\\
\nonumber
\bar{L}\gamma^{\mu}D^{\nu}LB_{\mu\nu}\\
\nonumber
\bar{\ell}_{R}\gamma^{\mu}D^{\nu}\ell_{R}B_{\mu\nu}\\
\label{eq:fermionhiggsgauge}
\bar{\nu}_{R}\gamma^{\mu}D^{\nu}\nu_{R}B_{\mu\nu}\\
\nonumber
g_{2}(\bar{L}\sigma^{\mu\nu}\tau^{a}\phi)\ell_{R}W_{\mu\nu}^{a}\\
\nonumber
g_{1}(\bar{L}\sigma^{\mu\nu}\phi)\ell_{R}B_{\mu\nu}\\
\nonumber
g_{2}(\bar{L}\sigma^{\mu\nu}\tau^{a}\widetilde{\phi})\nu_{R}W_{\mu\nu}^{a}\\
\nonumber
g_{1}(\bar{L}\sigma^{\mu\nu}\widetilde{\phi})\nu_{R}B_{\mu\nu}
\end{eqnarray}
As for the fermion-Higgs operators, the operators in (\ref{eq:fermionhiggsgauge}) that contain an even number of $\nu_R$ fields will not contribute significantly to $m_\nu^{AB}$, so only the last two in the list are relevant:
\begin{eqnarray}
{\cal O}_{B,\, AD}^{(6)} & = & g_{1}(\bar{L}^A\sigma^{\mu\nu}\widetilde{\phi})\nu_{R}^D B_{\mu\nu} \\
{\cal O}_{W,\, AD}^{(6)} & = & g_{2}(\bar{L}^A \sigma^{\mu\nu}\tau^{a}\widetilde{\phi})\nu_{R}^DW_{\mu\nu}^{a}
\end{eqnarray}

In addition to these operators, there exist additional $n=6$ operators that contain two derivatives. However, as discussed in Ref.~\cite{Bell:2005kz}, they can either be related to ${\cal O}_{B,\, AD}^{(6)}$
and ${\cal O}_{W,\, AD}^{(6)}$ through the equations of motion or contain derivatives acting on the $\nu_R$ fields so that they do not contribute to the neutrino mass operator. Consequently, we need not consider them here.  We also observe that the operator ${\cal O}^{(6)}_{W,\, AD}$ will also contribute to the $\mu$-decay  amplitude {\em via} graphs as in  Fig. \ref{fig:mutree}b. We have computed its contributions to the Michel parameters and find that they are suppressed by $(\sim m_\mu/\Lambda)^2\lsim 1.7\times 10^{-7}$ relative to the effects of the other $n=6$ operators. This suppression arises from  the presence of the derivative acting on the gauge field and the absence of an interference between the corresponding amplitude and that of the SM.  Finally, we note that the operators whose chiral structure suppresses their contributions to the neutrino mass operator (as discussed above) may, in general,  contribute to muon decay via the terms in Eq.~(\ref{eq:leff0}) having $\epsilon=\mu$. We do not consider these terms in this study.

\section{Operator Mixing}
\label{sec:mixing}

In analyzing mixing among operators that contribute to both $\mu$-decay and $m_\nu^{AD}$ it is useful to consider separately two cases: (i) mixing between the $n=6$ operators that enter $\mu$-decay and the $n=4$ mass operator, ${\cal O}^{(4)}_{M,\, AD}$, and (ii) mixing among the relevant $n=6$ operators. In general, contributions to $m_\nu^{AD}$ involving the second case will be smaller than those that involve mixing with ${\cal O}^{(4)}_{M,\, AD}$ by $\sim (v/\Lambda)^2$, since ${\cal O}^{(6)}_{M,\, AD}$ contains an additional factor of $(\phi^\dag\phi)/\Lambda^2$. We first consider this case and employ dimensional analysis to derive neutrino mass naturalness bounds on the $n=6$ operator coefficients. For $v$ not too different from $\Lambda$, the impact of the $n=6$ mixing can also be important, and in this case we can employ a full renormalization group (RG) analysis to derive robust naturalness bounds. 

\subsection{Mixing with ${\cal O}^{(4)}_{M,\, AD}$}

The analysis of Ref.~\cite{Prezeau:2004md} employed dimensional regularization (DR) to regularize the one- and two-loop graphs through which four-fermion operators containing a single $\nu_R$ field contribute to the $n=6$ mass operator. Mixing with lower-dimension operators cannot be treated using DR since the relevant graphs are quadratically divergent and must be proportional to the square of a mass scale. For $\mu>v$, all fields are massless, and $\mu$ itself appears only logarithmically. Since the mass operator exists for zero external momentum, all quadratically-divergent graphs vanish in this case. 

The $n=4$ mass operator will nevertheless receive contributions 
from the quadratically divergent graphs containing the $n=6$ operators. Since the integration region includes momenta of order the cut-off, these quadratically divergent graphs will have magnitude $\sim \Lambda^2/(4\pi)^2$. In DR, these contributions are absorbed into the operator coefficients $C_{M,\, AD}^4$ whose values are taken from experiment. One may, however, estimate the size of these contributions either using a gauge-invariant regulator, such as the generalized Pauli-Villars regulator of Ref. \cite{Frolov:1992ck}, or using dimensional arguments. Since we are interested in order-of-magnitude constraints, use of the latter is sufficient. 

\begin{figure}[h]
\centerline{\epsfig{figure=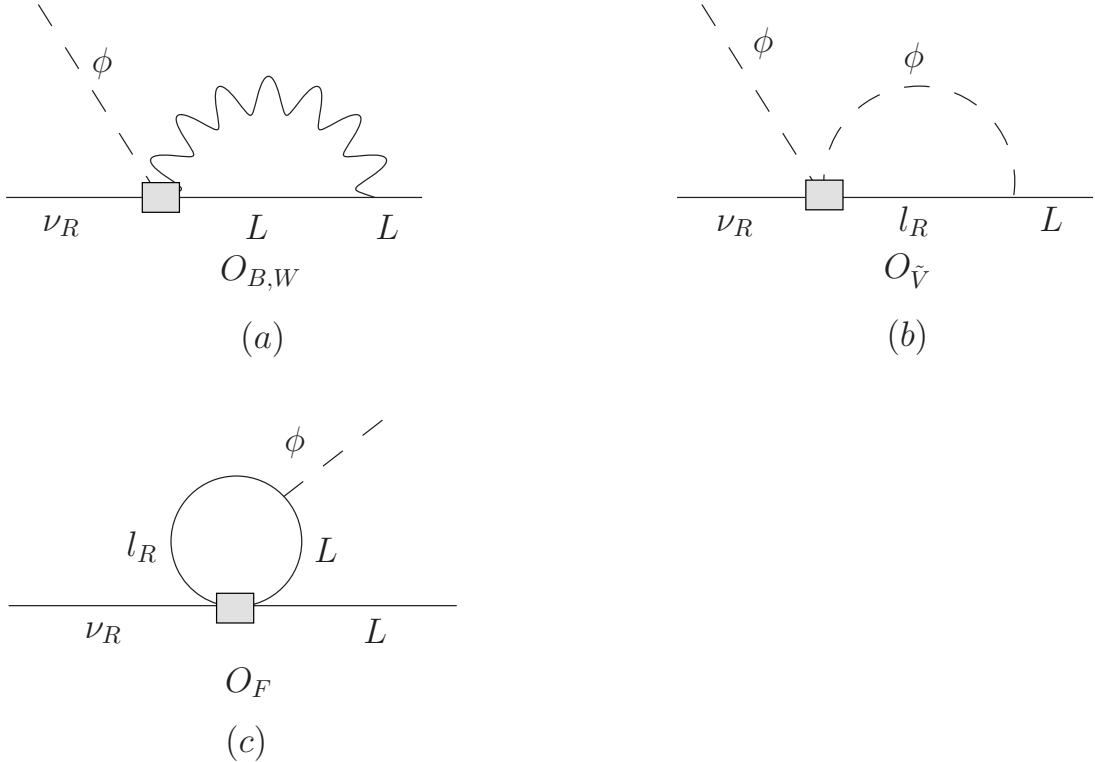,width=6.in}} \caption{{\em One loop graphs for the mixing of the $n=6$ operators (denoted by the shaded box) into the $n=4$ mass operator ${\cal O}^{(4)}_{M,\ AD}$ . Solid, dashed, and wavy lines denote fermions, Higgs scalars, and gauge bosons, respectively. Panels (a,b,c) illustrate mixing of ${\cal O}_{B,W}^{(6)}$, ${\cal O}^{(6)}_{{\tilde V}}$, and  ${\cal O}^{(6)}_{F}$, respectively, into ${\cal O}^{(4)}_{M,\ AD}$.}} \label{fig:neq4mix}
\end{figure}

The relevant one-loop graphs are shown in Fig. \ref{fig:neq4mix}. For the mixing of the four-fermion operators ${\cal O}^{(6)}_{F,\, ABCD}$ into ${\cal O}^{(4)}_{M,\, AD}$, two topologies are possible, associated with either the fields (${\bar L}^A$, $\nu_R^D$) or (${\bar L}^B$, $\nu_R^D$) living on the external lines. For the mixing of ${\cal O}^{(6)}_{F,\, ABCD}$ as well as of ${\cal O}^{(6)}_{{\tilde V},\, AB}$ into ${\cal O}^{(4)}_{M,\, AD}$, one insertion of the Yukawa interaction $f_{AC}^\ast\,  {\bar l}^C_R L^A$ is needed to convert the internal, RH lepton into a LH one. In contrast, no Yukawa insertion is required for the mixing of ${\cal O}_{B,\, AD}^{(6)}$ and ${\cal O}_{W,\, AD}^{(6)}$ into ${\cal O}^{(4)}_{M,\, AD}$.  

To simplify the analysis of mixing with the ${\cal O}^{(6)}_{F,\, ABCD}$ we note that one may always redefine the fields $L^A$ and $\ell_R^D$ so that the charged lepton Yukawa matrix $f_{AD}$ is diagonal. Specifically, we take
\begin{eqnarray}
\label{eq:redef1}
L^A & \rightarrow& L^{A\, \prime} = S_{AB} L^B \\
\nonumber
\ell_R^C  & \rightarrow& \ell^{C\, \prime} = T_{CD} \ell^D
\end{eqnarray}
with $S_{AB}$ and $T_{CD}$ chosen so that 
\begin{equation}
{\bar L}\, {\tilde f}\,  \ell = {\bar L}^\prime\, {\tilde f}_{\rm diag}\, \ell^\prime
\end{equation}
where $L$, $L^\prime$ denote vectors in flavor space, $\tilde f$ denotes the Yukawa matrix in the original basis, and  ${\tilde f}_{\rm diag} = {\tilde S}^\dag\, {\tilde f}\, {\tilde T}$. We note that the field redefinition (\ref{eq:redef1}) differs from the conventional flavor rotation used for quarks, since we have performed identical rotations on both isospin components of the left-handed doublet. Consequently, gauge interactions in the new basis entail no transitions between generations. We also note that Eqs. (\ref{eq:redef1}) also imply a redefinition of the operator coefficients $C^4_{M,\, AD}$, $C^6_{F,\, ABCD}$, {\em etc.}. For example, one has
\begin{eqnarray}
\label{eq:redef2}
C^{4,6}_{M,\, A^\prime D} & = & C^{4,6}_{M,\, AD}\, S_{M,\, A^\prime A}  \\
\nonumber
C^{6\, \prime}_{F,\, A^\prime B^\prime C^\prime D}& =& C^6_{F,\, ABCD}\,  S_{A^\prime A}\,  S_{B^\prime B}\, 
T^\ast_{C^\prime C} 
\end{eqnarray}
where a sum over repeated indices is implied. Diagonalization of the neutrino mass matrix requires additional, independent rotations of the $\nu_{L,R}^D$ fields after inclusion of radiative contributions to the coefficients $C^{4,6}_{M,\, AD}$ generated by physics above the weak scale. Since we are concerned only with contributions generated above the scale of SSB, we will not perform the latter diagonalization and carry out computations using the $L^\prime$, $\ell^\prime_R$ basis\footnote{ For notational simplicity, we henceforth omit the prime superscripts.}.  

In this case, the only four fermion operators ${\cal O}^{(6)}_{F,\, ABCD}$ that can contribute substantially to $m_\nu^{AD}$ are those having either $A=C$ or $B=C$. Thus, we obtain the following estimates of the contributions from the $n=6$ operators to the coefficient of the $n=4$ mass operator:
\begin{eqnarray}
\nonumber
{\cal O}_{B,\, AD}^{(6)} & \rightarrow & C^4_{M,\, AD} \sim \frac{\alpha}{4\pi \cos^2\theta_W} C^6_{B,\, AD} \\
\nonumber
{\cal O}_{W,\, AD}^{(6)} & \rightarrow & C^4_{M,\, AD} \sim \frac{3\alpha}{4\pi \sin^2\theta_W} C^6_{W,\, AD}\\
\label{eq:neq4}
{\cal O}^{(6)}_{{\tilde V},\, AD} & \rightarrow & C^4_{M,\, AD}\sim \frac{f_{AA}}{16\pi^2} C^6_{{\tilde V},\, AD}\\
\nonumber
{\cal O}^{(6)}_{F,\, ABAD} & \rightarrow & C^4_{M,\, BD}\sim \frac{f_{AA}}{4\pi^2} C^6_{F,\, ABAD}\\
\nonumber
{\cal O}^{(6)}_{F,\, ABBD} & \rightarrow & C^4_{M,\, AD}\sim \frac{f_{BB}}{16\pi^2} C^6_{F,\, ABBD}
\end{eqnarray}
where $\theta_W$ is the weak mixing angle. 

The relative factor of $3\cot^2\theta_W$ for the mixing of ${\cal O}_{W,\, AD}^{(6)}$ compared to the mixing of ${\cal O}_{B,\, AD}^{(6)}$ arises from the ratio of gauge couplings $(g/g')^2$ and the presence of a ${\vec\tau}\cdot{\vec\tau}$ appearing in Fig. \ref{fig:neq4mix}a. The factor of two that enters the mixing of  ${\cal O}^{(6)}_{F,\, ABAD}$ compared to that of ${\cal O}^{(6)}_{F,\, ABBD}$ arises from the trace associated with the closed chiral fermion loop that does not arise for ${\cal O}^{(6)}_{F,\, ABBD}$.

We observe that there exist two four-fermion operators that contribute to $\mu$-decay that do not contribute to $C^4_{M,\, AD}$ in the basis giving a diagonal $f_{AB}$: ${\cal O}^{(6)}_{F,\, AABD}$ with either $A=1,B=2$ or $A=2,B=1$. As we discuss in Section \ref{sec:mnuconstraints}, these operators contribute to $g^{S,T}_{LR}$ and $g^{S,T}_{RL}$, respectively. Consequently, the magnitudes of these couplings are not directly bounded by $m_\nu$ and naturalness considerations, as indicated in Table \ref{tab:gconstraints}. From a theoretical standpoint, one might expect the magnitudes of $C^6_{F,\, 112D}$ and $C^6_{F,\, 221D}$ to be comparable to those of the other four-fermion operator coefficients in models that are consistent with the scale of neutrino mass. Nevertheless, we cannot {\em a priori} rule out order of magnitude or more differences between operator coefficients. 

\subsection{Mixing among $n=6$ operators}

Because ${\cal O}^{(6)}_{M,\, AD}$ contains one power of $(\phi^\dag\phi)/\Lambda^2$ compared to ${\cal O}^{(4)}_{M,\, AD}$, the constraints obtained from mixing with the former will generally be weaker by $\sim (v/\Lambda)^2$. However, for $\Lambda$ not too different from the weak scale, the $n=6$ mixing can be of comparable importance to the $n=4$ case. Here, we study the mixing among $n=6$ operators by computing all one-loop graphs that contribute using DR and performing a renormalization group (RG) analysis. Doing so provides the exact result for contributions to the one-loop mixing from scales between $\Lambda$ and $v$, summed to all orders in $ f_{AA} \ln(v/\Lambda)$ and $\alpha \ln(v/\Lambda)$.

In carrying out this analysis, it is necessary to identify a basis of operators that close under renormalization. We find that the minimal set consists of seven operators that contribute to $\mu$-decay and $m_\nu^{AD}$:
\be
{\cal O}_{B,\, AD}^{(6)},\ {\cal O}_{W,\, AD}^{(6)},\ {\cal O}^{(6)}_{M,\, AD},\ {\cal O}^{(6)}_{{\tilde V},\, AD},\
{\cal O}^{(6)}_{F,\, AAAD},\ {\cal O}^{(6)}_{F,\, ABBD},\ {\cal O}^{(6)}_{F,\, BABD}\ \  \ .
\ee
For simplicity, we have included a single RH neutrino field $\nu_R^D$  in all seven operators. While one could, in principle, allow for different $\nu_R$ generation indices, the essential physics can be extracted from an analysis of this minimal basis. 

\begin{figure}[h]
\centerline{\epsfig{figure=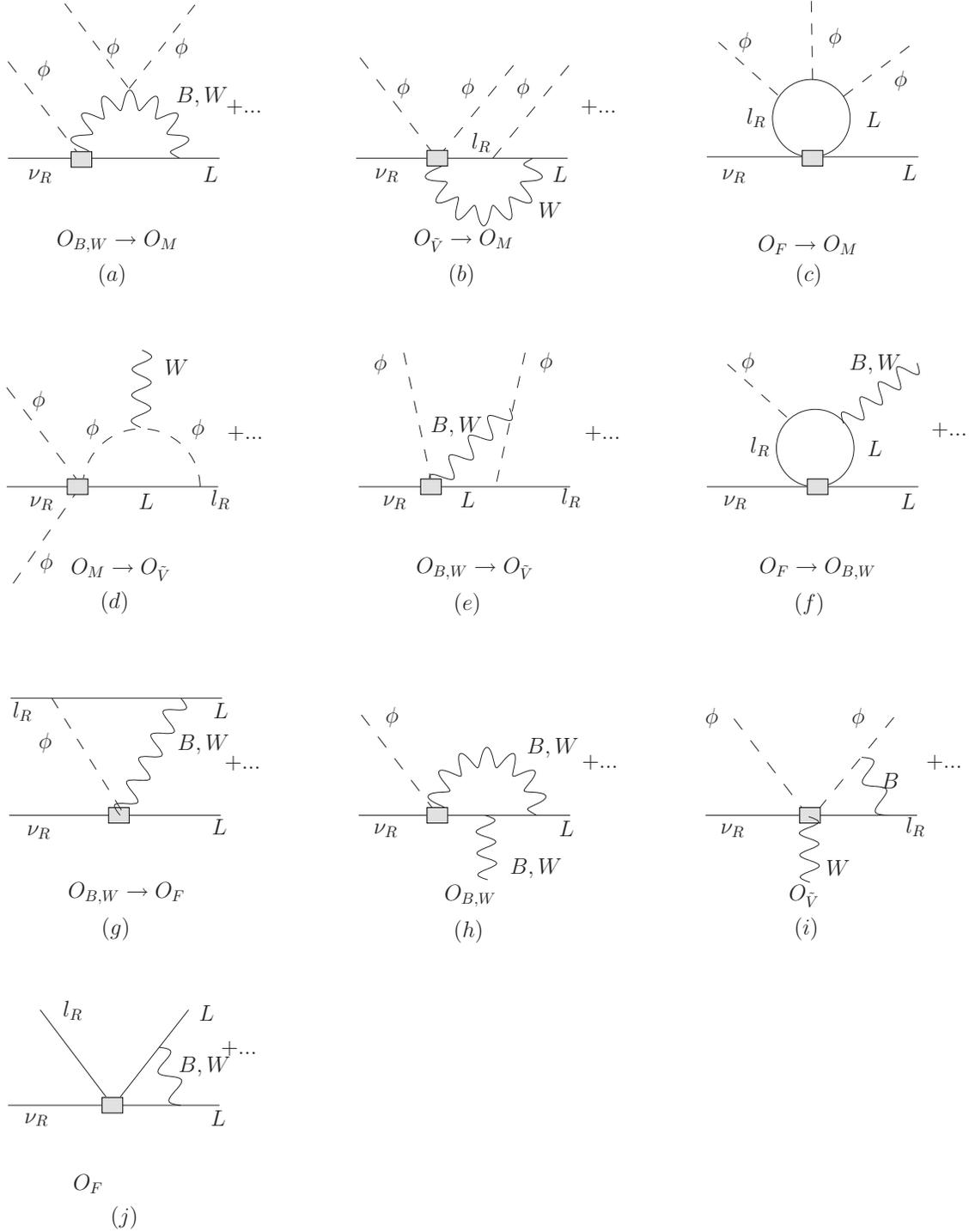,width=6.in}} \caption{{\em One loop graphs for the mixing among $n=6$ operators. Notation is as in previous figures. Various types of mixing (a-g) and self-renormalization (h-j) are as discussed in the text.}} \label{fig:neq6mix}
\end{figure}

The classes of graphs relevant to mixing among these operators are illustrated in Fig. \ref{fig:neq6mix}, where  we show representative contributions to operator self-renormalization and mixing among the various operators. The latter include mixing of all operators into ${\cal O}^{(6)}_{M,\, AD}$ (a-c); mixing of ${\cal O}^{(6)}_{M,\, AD}$, ${\cal O}_{B,\, AD}^{(6)}$, and ${\cal O}_{W,\, AD}^{(6)}$ into ${\cal O}^{(6)}_{{\tilde V},\, AD}$ (d,e); and mixing between the four-fermion operators and the magnetic moment operators (f,g). Representative self-renormalization graphs are given in Fig. \ref{fig:neq6mix}(h-j). As noted in Ref.~\cite{Prezeau:2004md}, the mixing of the the four-fermion operators into ${\cal O}^{(6)}_{M,\, AD}$ contains three powers of the lepton Yukawa couplings and is highly suppressed. In contrast, all other mixing contains at most one Yukawa insertion.

Working to first order in the $f_{AA}$ we find a total of 59 graphs that must be computed, not including wavefunction renormalization graphs that are not shown.  Twenty-two of these graphs were computed by the authors of Ref.~\cite{Bell:2005kz} in their analysis of the mixing between ${\cal O}^{(6)}_{M,\, AD}$ and the magnetic moment operators. Here, we compute the remaining 37. As in Ref.~\cite{Bell:2005kz}, we work with the background field gauge \cite{Abbott:1980hw} in $d=4-2\epsilon$ spacetime dimensions. We renormalize the operators using minimal subtraction, wherein counterterms simply remove the divergent, $1/\epsilon$ terms from the one-loop amplitudes. The resulting renormalized operators ${\cal O}^{(6)}_{jR}$ are expressed in terms of the unrenormalized operators ${\cal O}^{(6)}_{j}$ as
\be
{\cal O}^{(6)}_{jR} = \sum_k Z^{-1}_{jk} Z_L^{n_L/2} Z_\phi^{n_\phi/2} {\cal O}^{(6)}_{k} 
= \sum_k Z^{-1}_{jk} {\cal O}^{(6)}_{k0}\ \ \ ,
\ee
where
\be
{\cal O}^{(6)}_{j0}= Z_L^{n_L/2} Z_{\phi}^{n_\phi/2} {\cal O}^{(6)}_{j} 
\ee
are the $\mu$-independent bare operators, $Z_L^{1/2}$ and $Z_\phi^{1/2}$ are the wavefunction renormalization constants for the fields $L^A$ and $\phi$, respectively; $n_L$ and $n_\phi$ are the number of LH lepton and Higgs fields appearing in a given operator; and $Z^{-1}_{jk} Z_L^{n_L/2} Z_\phi^{n_\phi/2}$ are the counterterms that remove the $1/\epsilon$ divergences. 

Since the bare operators ${\cal O}^{(6)}_{j0}$ do not depend on the renormalization scale, whereas the $Z^{-1}_{jk}$ and  the ${\cal O}^{(6)}_{jR}$ do, the operator coefficients $C_j^6$ must carry a compensating $\mu$-dependence to ensure that ${\cal L}_{\rm eff}$ is independent of scale. This requirement leads to the RG equation for the operator coefficients:
\be 
\mu \frac{d}{d \mu}C^6_j +\sum_kC^6_k\ \gamma_{kj}=0
\ee
where 
\be
\gamma_{kj} = \sum_{\ell} \left( \mu\frac{d}{d\mu} Z_{k\ell }^{-1}\right)  Z_{\ell j} \ \ \ .
\ee
is the anomalous dimension matrix. We obtain\footnote{The term in $\gamma_{33}$ proportional to $\lambda$ differs from that of Ref.~\cite{Bell:2005kz}, which contains an error. However, this change does not  affect the bounds on the neutrino magnetic moments obtained in that work.} 
\begin{footnotesize}
\begin{eqnarray}
\label{eq:anomdim}
\gamma_{jk}=
\left( \begin{array}{ccccccc}
-\frac{3(\alpha_1-3\alpha_2)}{16\pi} & \frac{3\alpha_1}{8\pi} & -6\alpha_1(\alpha_1+\alpha_2) &
-\frac{9\alpha_1 f_{AA}^\ast}{8\pi}&  -\frac{9\alpha_1 f_{AA}}{4\pi} & -\frac{9\alpha_1 f_{BB}}{2\pi} &
\frac{9\alpha_1 f_{BB}}{4\pi}\\
\frac{9\alpha_2}{8\pi}& \frac{3(\alpha_1-3\alpha_2)}{16\pi}& 6\alpha_2(\alpha_1+3\alpha_2) &
\frac{27\alpha_2 f_{AA}^\ast}{8\pi} & -\frac{9\alpha_2 f_{AA}}{4\pi} & -\frac{9\alpha_2 f_{BB}}{2\pi} &
\frac{9\alpha_2 f_{BB}}{4\pi} \\
0 & 0 & \frac{9(\alpha_1+3\alpha_2)}{16\pi}-\frac{3\lambda}{2\pi^2} & 0 & 0 & 0 & 0 \\
0  & 0 & \frac{9\alpha_2 f_{AA}}{8\pi}-\frac{3 f_{AA}\lambda}{8\pi^2} & \frac{3\alpha_1}{4\pi} & 0 & 0 & 0   \\
-\frac{3 f_{AA}^\ast}{128\pi^2} & -\frac{ f_{AA}^\ast}{128\pi^2} & 0 & 0 &\frac{3(3\alpha_1-\alpha_2)}{8\pi} &
0 & 0 \\
-\frac{3 f_{BB}^\ast}{128\pi^2} & -\frac{ f_{BB}^\ast}{128\pi^2} & 0 & 0 &0 & \frac{3(\alpha_1+\alpha_2)}{8\pi} & \frac{3(\alpha_1-\alpha_2)}{4\pi}\\
0 & 0 & 0 & 0 & 0 & \frac{3(\alpha_1-\alpha_2)}{4\pi} & \frac{3(\alpha_1+\alpha_2)}{8\pi} 
\end{array} \right)
\end{eqnarray}
\end{footnotesize}
where the $\alpha_i=g_i^2/(4\pi)$ and $\lambda$ is the Higgs self coupling defined by the potential $V(\phi)=\lambda[(\phi^\dag\phi)-v^2/2]^2$.


\begin{figure}[h]
\centerline{\epsfig{figure=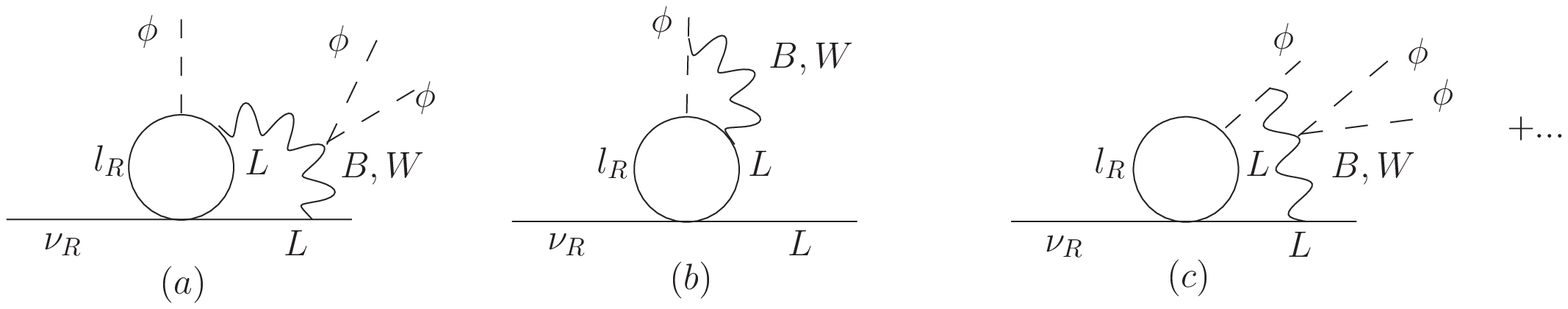,width=6.in}} \caption{{\em Two-loop graphs for the mixing of the $n=6$ operators. Only representive graphs for the mixing of the four-fermion operators ${\cal O}^{(6)}_{F,\, ABCD}$ into ${\cal O}^{(6)}_{M,\, AD}$ are shown. }} \label{fig:neq6twoloop}
\end{figure}

Using this result for $\gamma_{ij}$ and the one-loop $\beta$ functions for $\alpha_1$, $\alpha_2$, and the lepton Yukawa couplings, we solve the RG equations to determine the operator coefficients $C_k^6(\mu)$ as a function of their values at the scale $\Lambda$. As in Ref.~\cite{Bell:2005kz} we find that the the running of the gauge and Yukawa couplings has a negligible impact on the evolution of the 
$C_k^6(\mu)$. It is instructive to consider the results obtained by retaining only the leading logarithms $\ln(\mu/\Lambda)$ and terms at most first order in the Yukawa couplings. We find
\begin{eqnarray}
\nonumber
C^6_{M,\, AD}(\mu) & = & C^6_{M,\, AD}(\Lambda) \left[1-\gamma_{33}\ln\frac{\mu}{\Lambda}\right]\\
\nonumber
	&&-\left[\gamma_{-} C^6_{-} (\Lambda) +\gamma_{+}C^6_{+}(\Lambda) +\gamma_{43}C^6_{{\tilde V},\, AD}(\Lambda)\right]\ln\frac{\mu}{\Lambda}\\
\nonumber
C^6_{+}(\mu) & = & C^6_{+}(\Lambda)\left[1-{\tilde\gamma}\ln\frac{\mu}{\Lambda}\right]\\
\nonumber
&&+\left[\left(f_{AA}^\ast/32\pi^2\right)C^6_{F,\ AAAD}(\Lambda)+\left(
f_{BB}^\ast/32\pi^{2}\right)  C_{F,\ ABBD}^{6}(\Lambda) \right]\ln\frac{\mu}{\Lambda}\\
\nonumber
{\tilde C}^6(\mu) & = & {\tilde C}^6(\Lambda)\left[1+{\tilde\gamma}\ln\frac{\mu}{\Lambda}\right] \\
\nonumber
&& +[\left(3f_{AA}/128\pi^2\right)\left(\alpha_1-\alpha_2\right)C^6_{F,\ AAAD}(\Lambda)\\ 
\nonumber
&& +\left(  3f_{BB}/128\pi^{2}\right)  \left(\alpha_{1}-\alpha_{2}\right)  C_{F,\ ABBD}^{6}(\Lambda)]\ln\frac{\mu}{\Lambda}\\
\label{eq:oneloop}
C^6_{{\tilde V},\, AD}(\mu) & = &C^6_{{\tilde V},\, AD}(\Lambda) \left[1-\gamma_{44}\ln\frac{\mu}{\Lambda}\right]+(9f_{AA}/8\pi){\tilde C}^{6}(\Lambda)\ln\frac{\mu}{\Lambda}\\
\nonumber
C_{F,\ AAAD}^{6}(\mu)& =&C_{F,\ AAAD}^{6}(\Lambda)\left[1+\frac{3(\alpha_{2}-3\alpha_{1})}{8\pi}\ln\frac{\mu}{\Lambda}\right]\\
\nonumber
&& +(9f_{AA}/4\pi)\left[  C_{B,\,AD}^{6}(\Lambda)\alpha_{1}
+C_{W,\,AD}^{6}(\Lambda)\alpha_{2}\right]  \ln\frac{\mu}{\Lambda}\\
\nonumber
C_{F,\ ABBD}^{6}(\mu)& =&C_{F,\ ABBD}^{6}(\Lambda)\left[1-\frac{3(\alpha_{1}+\alpha_{2})}{8\pi}\ln\frac{\mu}{\Lambda}\right]\\
\nonumber
&& -\frac{3(\alpha_{1}-\alpha_{2})}{4\pi}C_{F,\ BABD}^{6}(\Lambda)\ln\frac{\mu}{\Lambda}\\
\nonumber
&& +(9f_{BB}/2\pi)\left[  C_{B,\,AD}^{6}(\Lambda)\alpha_{1}+C_{W,\,AD}^{6}(\Lambda)\alpha_{2}\right]  \ln\frac{\mu}{\Lambda}\\
\nonumber
C_{F,\ BABD}^{6}(\mu)  & =&C_{F,\ BABD}^{6}(\Lambda)\left[1-\frac{3(\alpha_{1}+\alpha_{2})}{8\pi}\ln\frac{\mu}{\Lambda}\right]\\
\nonumber
&& -\frac{3(\alpha_{1}-\alpha_{2})}{4\pi}C_{F,\ ABBD}^{6}(\Lambda)\ln\frac{\mu}{\Lambda}\\
\nonumber
&& -(9f_{BB}/4\pi)\left[C_{B,\,AD}^{6}(\Lambda)\alpha_{1}+C_{W,\,AD}^{6}(\Lambda)\alpha_{2}\right]  \ln\frac{\mu}{\Lambda}%
\end{eqnarray}
where 
\begin{eqnarray}
\nonumber
C^6_{\pm}(\mu) & \equiv& C^6_{B,\, AD}(\mu) \pm C^6_{W,\, AD}(\mu) \\
{\tilde C}^6(\mu) & \equiv & \alpha_1 C^6_{B,\ AD}(\mu)-3\alpha_2 C^6_{W,\, AD}(\mu) \\
\nonumber
\gamma_{\pm} & \equiv& \left(\gamma_{13}\pm \gamma_{23}\right)/2 \\
\nonumber
\tilde\gamma& \equiv & 3(\alpha_1+3\alpha_2)/16\pi
\end{eqnarray}

We note that the combination of coefficients $C^6_{+}(v)$ enters the neutrino magnetic moment. Its RG evolution was obtained in Ref.~\cite{Bell:2005kz} to zeroth order in the Yukawa couplings; here we obtain the corrections that are linear in $f_{AA}$ and $f_{BB}$. The corresponding contributions to the neutrino mass matrix $\delta m_\nu^{AD}$ and magnetic moment matrix $\mu_\nu^{AD}$ are then given by
\begin{eqnarray}
\label{eq:msix}
\delta m_\nu^{AD} & = & -\left(\frac{v^3}{2\sqrt{2}\Lambda^2}\right) C^6_{M,\, AD}(v)\\
\label{eq:musix}
\frac{\mu_\nu^{AD}}{\mu_B} & = & -4\sqrt{2}\left(\frac{m_e v}{\Lambda^2}\right)\, {\rm Re}\left\{ C^6_{+}(v)\right\} \ \ \ .
\end{eqnarray}

From Eqs. (\ref{eq:oneloop},\ref{eq:msix},\ref{eq:musix}) we observe that to linear order in the lepton Yukawa couplings, $C^6_{M,\, AD}(\mu)$ receives contributions from the two magnetic moment operators and ${\cal O}^{(6)}_{{\tilde V}}$ but not from the four fermion operators. This result is consistent with the result obtained by the authors of Ref.~\cite{Prezeau:2004md}, who computed one-loop graphs containing the four-fermion operators of Eq. (\ref{eq:leff0}) using massive charged leptons and found that contributions to $m_\nu\propto m_\ell^3$. In the effective theory used here, the latter result corresponds to a one-loop computation with three insertions of the Yukawa interaction. However, mixing with ${\cal O}^{(6)}_{\tilde V}$ was not considered in Ref.~\cite{Prezeau:2004md}, and our result that this operator mixes with ${\cal O}^{(6)}_{M,\, AD}$ to linear order in the Yukawa couplings represents an important difference with the former analysis. 

We agree with the observation of Ref.~\cite{Prezeau:2004md} that the four fermion operators can mix with ${\cal O}^{(6)}_{M,\, AD}$ to linear order in the $f_{AA}$ via two-loop graphs, such as those indicated in Fig. \ref{fig:neq6twoloop}. These graphs were estimated in Ref.~\cite{Prezeau:2004md} by considering loops with massive $W^\pm$ and $Z^0$ bosons that correspond in our framework to the diagrams of Fig. \ref{fig:neq6twoloop}a. We observe, however, that the two-loop constraints will be weaker than those obtained by one-loop mixing with ${\cal O}^{(4)}_{M,\, AD}$ by $\sim (\alpha/4\pi)(v/\Lambda)^2$ (modulo logarithmic corrections), so we do not consider this two-loop mixing in detail here. Moreover, because we work at a scale $\mu> v$ for which the use of massless fields is appropriate, and because we adopt a basis in which the Yukawa matrix and gauge interactions are flavor diagonal (but $m_\nu^{AD}$ is not), the operators ${\cal O}^{(6)}_{F,\, 112D}$ and
${\cal O}^{(6)}_{F,\, 221D}$ will not mix with  ${\cal O}^{(6)}_{M,\, AD}$ even at two-loop order.

\section{Neutrino Mass Constraints}
\label{sec:mnuconstraints}

To arrive at neutrino mass naturalness constraints on the $g^\gamma_{\epsilon\mu}$ coefficients, it is useful to tabulate their relationships with the dimension six operator coefficients. In some cases, one must perform a Fierz transformation in order to obtain the operator structures in Eq. (\ref{eq:leff0}). Letting
\begin{equation}
\label{eq:gcrel}
g^\gamma_{\epsilon\mu} = -\kappa \left(\frac{v}{\Lambda}\right)^2 C^6_k
\end{equation}
we give in Table \ref{tab:gcrel}  the $\kappa$s corresponding to the various dimension six operators. 

Using the entries in Table \ref{tab:gcrel} and the estimates in Eqs. (\ref{eq:neq4}), we illustrate how the bounds in Table \ref{tab:gconstraints} were obtained. For the operator ${\cal O}^{(6)}_{F,\, 122D}$, for example, we have from Eqs. (\ref{eq:neq4mass},\ref{eq:neq4})
\begin{equation}
|C^6_{F,\, 122D}| \lsim 16\pi^2 \left(\frac{\delta m_\nu^{1D}}{m_\mu}\right)
\end{equation}
leading to 
\begin{equation}
|g^S_{LR}| \lsim 4\pi^2 \left(\frac{\delta m_\nu^{1D}}{m_\mu}\right) \left(\frac{v}{\Lambda}\right)^2 \qquad
|g^T_{LR}| \lsim 2\pi^2 \left(\frac{\delta m_\nu^{1D}}{m_\mu}\right) \left(\frac{v}{\Lambda}\right)^2 
\end{equation}
where $\delta m_\nu^{AD}$ denotes the radiative contribution to $m_\nu^{AD}$. Choosing $\Lambda=v$ and $\delta m_\nu^{1D} = 1$eV (corresponding to the scale of upper bounds derived from $^3$H $\beta$-decay studies\cite{Lobashev:1999tp,Weinheimer:1999tn}) leads to the bounds in the first row of Table \ref{tab:gconstraints}. Similar arguments yield the other entries in the table. Note that the bounds become smaller as $\Lambda$ is increased from $v$. 

The constraints on the $g^V_{LR,RL}$ that follow from mixing among the $n=6$ operators follows straightforwardly from Eqs. (\ref{eq:oneloop}, \ref{eq:msix}) and Table \ref{tab:gcrel}. We obtain
\begin{equation}
\label{eq:gvsix}
g^V_{LR} = \left(\frac{\delta m_\nu^{2D}}{m_\mu}\right) \left(\frac{8\pi\sin^2\theta_W}{9}\right)
\left(\alpha-\frac{\lambda\sin^2\theta_W}{3\pi}\right)^{-1}\left(\ln\frac{v}{\Lambda}\right)^{-1}\ \ \ .
\end{equation}
A similar expression holds for $g^V_{RL}$ but with $m_\mu\to m_e$ and $\delta m_\nu^{2D}\to\delta m_\nu^{1D}$. Note that in arriving at Eq. (\ref{eq:gvsix}) we have ignored the running of the $C^6_{{\tilde V},\, AD}(\mu)$ between $\Lambda$ and $v$, since the impact on the $g^V_{LR,RL}$ is higher order in the gauge and Yukawa couplings. To derive numerical bounds on the $g^V_{LR,RL}$ from Eq. (\ref{eq:gvsix}) we use the running couplings in the $\overline{\rm MS}$ scheme $\alpha={\hat\alpha}(M_Z)\approx 1/127.9$, $\sin^2{\hat\theta}_W(M_Z) \approx 0.2312$ and the tree-level relation between the Higgs quartic coupling $\lambda$, the Higgs mass $m_H$, and $v$: $2\lambda=(m_H/v)^2$. We quote two results, corresponding to the direct search lower bound on $m_H\gsim 114$ GeV and the one-sided 95 \% C.L. upper bound from analysis of precision electroweak measurements, $m_H\lsim 186$ GeV\cite{LEP}. We obtain
\begin{eqnarray}
\label{eq:gvsixb}
\left\vert g^V_{LR}\right\vert  & =&   \left(\frac{\delta m_\nu^{2D}}{1\, {\rm eV}}\right) \left(\ln\frac{\Lambda}{v}\right)^{-1}\, 
\begin{cases}
1.2\times 10^{-6}, & m_H=114\, {\rm GeV}\\
7.5\times 10^{-6}, & m_H=186\, {\rm GeV}
\end{cases} \\
\nonumber
\left\vert g^V_{RL}\right\vert  & =&   \left(\frac{\delta m_\nu^{1D}}{1\, {\rm eV}}\right) \left(\ln\frac{\Lambda}{v}\right)^{-1}\, 
\begin{cases}
2.5 \times 10^{-4}, & m_H=114\, {\rm GeV}\\
1.5\times 10^{-3}, & m_H=186\, {\rm GeV}
\end{cases}
\end{eqnarray}
For $\Lambda\sim 1$ TeV, the logarithms are ${\cal O}(1)$ so that for $\delta m_\nu\sim 1$ eV, the bounds on the $g^V_{LR,RL}$ derived from $n=6$ mixing are comparable in magnitude to those estimated from mixing with the $n=4$ mass operators. 

Although the four fermion operators do not mix with ${\cal O}^{(6)}_{M,\, AD}$ at linear order in the Yukawa couplings, they do contribute to the magnetic moment operators ${\cal O}_{B,\, AD}^{(6)}$ and $\ {\cal O}_{W,\, AD}^{(6)}$ at this order. From Eqs.(\ref{eq:oneloop}, \ref{eq:musix}) we have
\begin{equation}
\label{eq:mu4f}
\frac{\delta \mu_\nu^{AD}}{\mu_B} =  \frac{\sqrt{2}}{8\pi^2}\, \left(\frac{m_e}{v}\right)\left(\frac{v}{\Lambda}\right)^2\, {\rm Re}\, 
\left[f_{AA}^\ast C^6_{F,\, AAAD}+f_{BB}^\ast C^6_{F,\, ABBD}\right]\, \ln\frac{\Lambda}{v}\ \ \ ,
\end{equation}
where $\delta\mu_\nu^{AD}$ denotes the contribution to the magnetic moment matrix and $\mu_B$ is a Bohr magneton. While  ${\cal O}^{(6)}_{F,\, AAAD}$ does not contribute to $\mu$-decay, the operator
${\cal O}^{(6)}_{F,\, ABBD}$ does, and its presence in Eq.~(\ref{eq:mu4f}) implies constraints on its coefficient from current bounds on neutrino magnetic moments. The most stringent constraints arise for $A=1$, $B=2$ for which we find
\begin{equation}
\label{eq:mubounds}
|C^6_{F,\, 122D}| \left(\frac{v}{\Lambda}\right)^2 \lsim 5\times 10^{10} \left(\ln\frac{\Lambda}{v}\right)^{-1} \left( \frac{ \mu_\nu^{1D}}{\mu_B} \right) \ \ \ .
\end{equation}
Current experimental bounds on $|\mu_\nu^{\rm exp}/{\mu_B}|$ range from $\sim 10^{-10}$ from observations of solar and reactor neutrinos\cite{Beacom:1999wx,Liu:2004ny,Daraktchieva:2005kn,Xin:2005ky} to $\sim 3\times 10^{-12}$ from the non-observation of plasmon decay into ${\bar\nu}\nu$ in astrophysical objects\cite{Raffelt:1999gv}. Assuming that the logarithm in Eq.~(\ref{eq:mubounds}) is of order unity, these limits translate into bounds on $g^S_{LR}$ and $g^T_{LR}$ ranging from $\sim 1\to 0.03$ and $\sim 0.3\to 0.01$, respectively. The solar and reactor neutrino limits on $|\mu_\nu^{\rm exp}/{\mu_B}|$ imply bounds on the $g^{S,T}_{LR}$ that are weaker than those obtained from the global analysis of $\mu$-decay measurements, while those associated with the astrophysical magnetic moment limits are comparable to the global values.  Nevertheless, the bounds derived from neutrino magnetic moments are several orders of magnitude weaker than those derived from the scale of neutrino mass.

\begin{table}
\caption{Coefficients $\kappa$ that relate $g^\gamma_{\epsilon\mu}$ to the dimension six operator coefficients $C^6_k$ via Eq. (\ref{eq:gcrel}).}
\label{tab:gcrel}
\begin{tabular}{@{}lcccccc @{}}
\\
$\kappa$ & $g^S_{LR}$  & $g^T_{LR}$ & $g^S_{RL}$ & $g^T_{RL}$ & $g^V_{LR}$ & $g^V_{RL}$
\\
\hline
\\
$C^6_{F,\, 122D}$ & $1/4$ & $1/8$ & - & - & - & - \\ 
$C^6_{F,\, 212D}$ & $1/2$ & - & - & - & - & - \\ 
$C^6_{F,\, 112D}$ & $3/4$ & $1/8$  & - & - & - & - \\ 
$C^6_{F,\, 211D}$  & - & -&  $1/4$ & $1/8$ & - & - \\ 
$C^6_{F,\, 121D}$ & - & - & $1/2$ & - & - & - \\ 
$C^6_{F,\, 221D}$ & - & - &  $3/4$ & $1/8$ & - & - \\
$C^6_{{\tilde V},\, 2D}$ & - & - & - & - & $-1/2$ & - \\ 
$C^6_{{\tilde V},\, 1D}$ & - & - & - & - & - & $-1/2$  \\ \\ 
\hline
\end{tabular}
\end{table}

The naturalness bounds on the $C^6_k$ associated with the scale of $m_\nu$ have implications for the interpretation of $\mu$-decay experiments. Because the coefficients $C^6_{F,\, 112D}$ and $C^6_{F,\, 221D}$ that contribute to $g^{S,T}_{LR,RL}$ are not directly constrained by $m_\nu$, none of the eleven Michel parameters is directly constrained by neutrino mass alone. Instead, it is more relevant to compare the results of global analyses from which limits on the $g^\gamma_{\epsilon\mu}$ are obtained with the $m_\nu$ naturalness bounds, since the latter imply tiny values for the couplings $g^V_{LR,RL}$. Should future experiments yield a value for either of these couplings that is considerably larger than our bounds in Table \ref{tab:gconstraints}, the new physics above $\Lambda$ would have to exhibit either fine-tuning or a symmetry in order to evade unacceptably large contributions to $m_\nu$. In addition, should future global analyses find evidence for non-zero $g^{S,T}_{LR,RL}$ with magnitudes considerably larger than given by the $m_\nu$-constrained contributions listed in Table \ref{tab:gconstraints}, then one would have evidence for a non-trivial flavor structure in the new physics that allows considerably larger effects from the operators ${\cal O}^{(6)}_{F,\, 112D}$  and ${\cal O}^{(6)}_{F,\, 221D}$ than from the other four fermion operators. 

Finally, we note that one may use a combination of neutrino mass and direct studies of the Michel spectrum to derive bounds on a subset of the Michel parameters that are more stringent than one obtains from $\mu$-decay experiments alone. To illustrate, we consider the parameters $\delta$ and $\alpha$, for which one has
\begin{eqnarray}
\frac{3}{4}-\rho & = & \frac{3}{4}\left\vert g^V_{LR}\right\vert^2 + \frac{3}{2}\left\vert g^T_{LR}\right\vert^2
+\frac{3}{4} {\rm Re}\left(g^S_{LR} g^{T\, \ast}_{LR}\right) +\left(L\leftrightarrow R\right)\\
\alpha & = & 8\, {\rm Re}\left\{ g^V_{RL}\left(g^{S\, \ast}_{LR}+6 g^{T\, \ast}_{LR}\right)+\left(L\leftrightarrow R\right)\right\}\ \ \ .
\end{eqnarray}
From Table \ref{tab:gconstraints}, we observe that the magnitudes of the $g^V_{LR,RL}$ contributions to $\rho$ and $\alpha$ are constrained to be several orders of magnitude below the current experimental sensitivities, whereas the contributions $g^{S,T}_{LR,RL}$ that arise from ${\cal O}^{(6)}_{F,\, 112D}$ and ${\cal O}^{(6)}_{F,\, 221D}$ are only directly constrained by experiment. Thus, we may use the current experimental results for $\rho$ to constrain the operator coefficients  $C^6_{F,\, 112D}$ and $C^6_{F,\, 221D}$ and subsequently employ the results -- together with the $m_\nu$ bounds on the $g^V_{LR,RL}$ -- to derive expectations for the magnitude of $\alpha$. For simplicity, we consider only the contributions from $C^6_{F,\, 112D}$ to $\rho$, and using the current experimental uncertainty in this parameter, we find
\begin{equation}
\label{eq:c6Fbound}
\left\vert C^6_{F,\, 112D}\right\vert \left(\frac{v}{\Lambda}\right)^2 \lsim 0.1\ \ \ .
\end{equation}
In the parameter $\alpha$, this coefficient interferes with $C^6_{{\tilde V},\, 1D}$:
\begin{equation}
\alpha = -6 \left(\frac{v}{\Lambda}\right)^4\, {\rm Re}\left(C^6_{{\tilde V},\, 1D} C^{6\, \ast}_{F,\, 112D}+\cdots\right)\ \ \ ,
\end{equation}
where the \lq\lq $+\cdots$ " indicate contributions from the other coefficients that we will assume to be zero for purposes of this discussion. From Eq.~(\ref{eq:c6Fbound}) and the $m_\nu$ limits on $C^6_{{\tilde V},\, 1D}$ we obtain
\begin{equation}
\label{eq:alphabound}
\left\vert\alpha\right\vert \lsim 2\times 10^{-4}\, \left(\frac{v}{\Lambda}\right)^2\,  \left(\frac{m_\nu^{1D}}{1\, {\rm eV}}\right)\ \ \ .
\end{equation}
For $\Lambda=v$, this expectation for $|\alpha|$ is more than two orders of magnitude below the present experimental sensitivity and will fall rapidly as $\Lambda$ increases from $v$. A similar line of reasoning can be used to constrain the parameter $\alpha^\prime$ in terms of $m_\nu$ and the CP-violating phases that may enter the effective operator coefficients.

\section{Conclusions}
\label{sec:conclusions}

The existence of the small, non-zero masses of neutrinos have provided our first direct evidence for physics beyond the minimal Standard Model, and the incorporation of $m_\nu$ into SM extensions is a key element of beyond-the-SM model building. At the same time, the existence of non-vanishing neutrino mass -- together with its scale -- have important consequences for the properties of neutrinos and their interactions that can be delineated in a model-independent manner~\cite{Bell:2005kz,Davidson:2005cs,Prezeau:2004md,Ito:2004sh}. In this paper, we have analyzed those implications for the decay of muons, using the effective field theory approach of Ref.~\cite{Bell:2005kz} and concentrating on the case of Dirac neutrinos. We have derived model-independent naturalness bounds on the contributions  to the Michel parameters from various $n=6$ operators that also contribute to the neutrino mass matrix {\em via} radiative corrections. 

Our work has been motivated by the ideas in Ref.~\cite{Prezeau:2004md}, but our conclusions differ in important respects. Importantly, we find --- after properly taking into account SU(2)$_L\times$U(1)$_Y$ gauge invariance and mixing between $n=6$ and $n=4$ operators that cannot be studied using dimensional regularization --- the dominant constraints on the contributions to the Michel parameters occur at one-loop order, rather than through two-loop effects as in Ref.~\cite{Prezeau:2004md}. Consequently, the bounds we derive are generally two orders of magnitude (or more) stronger than those of Ref.~\cite{Prezeau:2004md}. In addition, we carefully study the flavor structure of the operators that can contribute to $\mu$-decay and and find that there exist four-fermion operators that do not contribute to the neutrino mass matrix through radiative corrections. These operators contribute to the effective scalar and tensor couplings $g^{S,T}_{LR,RL}$ of Eq.~(\ref{eq:leff0}). In contrast, all operators that generate the $g^V_{LR,RL}$ terms contribute to $m_\nu^{AD}$,  so these effective couplings do have neutrino-mass naturalness bounds. From a model-building perspective it might seem reasonable to expect the coefficients of the unconstrained four-fermion operator coefficients to have the same magnitude as those that are constrained by $m_\nu$, but is important for precise muon-decay experiments to test this expectation.

While we have focused on the implications of Dirac mass terms,  a similar analysis for the Majorana neutrinos is clearly called for. Indeed, in the case of neutrino magnetic moments, the requirement of flavor non-diagonality for Majorana magnetic moments leads to substantially weaker naturalness bounds than for Dirac moments. While we do not anticipate similar differences between the Majorana and Dirac case for operators that contribute to $\mu$-decay, a detailed comparison will appear in a forthcoming publication.

\acknowledgments
The authors thank N. Bell, V. Cirigliano, M. Gorshteyn, P. Vogel, and M. Wise for helpful discussions. This work was supported in part under U.S. Department of Energy contracts FG02-05ER41361 and DE-FG03-ER40701 and National Science Foundation award PHY-0071856.



\end{document}